# Lures of Engagement: An Outlook on Tactical AI Art

**Dejan Grba**
dejangrba@gmail.com
Artist, researcher, and scholar, Belgrade, Serbia

This paper aims to diversify the existing critical discourse by introducing new perspectives on the poetic, expressive, and ethical features of tactical media art that involves artificial intelligence (AI). It explores diverse artistic approaches and their effectiveness in critiquing the epistemic, phenomenological, and political aspects of AI science and technology. Focusing on the three representative thematic areas—sociocultural, existential, and political—it discusses works that exemplify poetic complexity and manifest the ambiguities indicative of a broader milieu of contemporary art, culture, economy, and society. With a closing summary of the major issues and possible directions to address them, the paper shows that tactical AI art provides important insights into the AI-influenced world and has the potential to advance computational arts toward a socially responsible and epistemologically relevant expressive stratum.

## 1. Tactical AI Art

Since its largely obscure beginnings in the 1970s (Wilson 2002), AI art has expanded, gained visibility, and attained sociocultural relevance since the second half of the 2010s (Burbano and West 2020). This was facilitated by the accelerating affordance of multilayered subsymbolic machine learning (ML) architectures such as Deep Learning (DL), and by the raising sociopolitical impact of AI technologies. Contemporary AI art includes diverse creative approaches to, and various degrees of technical involvement with, ML (Grba 2022). Its topics, methodologies, presentational formats, and implications are closely related to a range of disciplines in AI research, development, and application. AI art is affected by the epistemic uncertainties, conceptual challenges, conflicted paradigms, discursive issues, ethical, and sociopolitical problems of AI science and industry. Similar to other new media art disciplines, AI art has had an ambivalent relationship with the mainstream contemporary artworld (MCA), marked by selective marginalization and occasional exploitation (Bishop 2012; Grba 2021, 252-254).

Informed by the functions, applications, consequences, and other aspects of modern ML systems, AI artists have been increasingly engaging in the critique of the epistemological, existential, or sociopolitical issues of applied AI (Grba 2022, 3-17). Their production continues the heterogeneous flux of tactical media practices that have energized art and culture since the late 20[th] century with hybrid forms of academic criticism of, or critical interventions into, technological, political, economic, and cultural layers of the neoliberal condition. With the raising accessibility of technologies that can be modified and repurposed by the actors who operate outside of the established hierarchies of power and knowledge, tactical media has emerged as a response to a shift in the nature of power in postindustrial society toward the information economy in which efficiency, operationalism, and instrumental rationality become core values, and market transactions the predominant social good.

In different ways and contexts, tactical media artists subvert and expose the exploitative corporate strategies based on quantization, statistical reductionism, data-mining, behavioural tracking, prediction, and manipulation of decision-making (Grba 2020, 71-73). Their expressive forms do not always clearly match the explicit activist category but offer subtle, sometimes covert, critique. Tactical media works are not sweeping revolutionary events but engage in a micro-politics of disruption, intervention, and education. The adjective "tactical" also indicates that absolute victory and fundamental structural transformation are neither desirable nor truly attainable objectives; tactical media projects are fleeting, ephemeral, and pliable and their statements and actions must be continually reconfigured in response to their changing targets. Although it often maps the top-down power relations, tactical media embodies a sense of bottom-up resistance in a manner and style associated with cultural dissent and opposition. It challenges the dominant semiotic regime through signs, messages, and narratives that foster critical thinking. It offers new ways of seeing, understanding, and (in some scenarios) interacting with the targeted systems of power. The transformative effects of tactical media projects are often not immediate but cumulative and relational because they provide insights and tools that may become transformative in the hands of the audience. Successful works emphasize the audience's presence, experience, engagement, and response (Bourriaud 2002). Sometimes, however, the intentionally constrained audience engagement can also be worthy: the restricted interaction with the work helps us become aware of our limitations to make an immediately perceptible impact on the socioeconomic and political systems represented by the work (Raley 2009).

Tactical media is often so entangled with its core informational and technological apparatuses that protest in a sense becomes the mirror image of its target. As Raley puts it (2009, 30), "while [these] critical practices do not have the hollowness or emptiness of Space Invaders [game]—the paradigmatic scene of the individual fighting back against a relentless and formless enemy—at times they participate in the same solitary, and sedentary, aesthetic." Furthermore, by openly identifying loopholes and weaknesses in the systems they critique, tactical media artists set their efforts and achievements up for recuperation and exploitation (Lovink and Rossiter 2005). In some instances, tactical media even lapses into its opposite and becomes the sophisticated mystification of the California Ideology whereby a technocratic class of avant-garde artisans acts on behalf of "the [lay] people" by articulating a vision of individual freedom realizable from within the power structures of the information society (Barbrook and Cameron 1995/2008).

Tactical AI art inherits, extends, and often amplifies, the strengths and shortcomings of tactical media. This makes it conducive, both explicitly and implicitly, to understanding how contemporary AI reflects, influences, and produces sociopolitical relations, economies, and worldviews. The existing literature pertinent for tactical AI art includes Marcus and Davis' *Rebooting AI* (2019) as well as Mitchell's *Artificial Intelligence* (2019) which provide the conceptual, technological, and sociocultural critiques of AI research and implementation; Pasquinelli's *How a Machine Learns and Fails* (2019) and Kearns and Roth's *The Ethical Algorithm* (2019) that address the ethical, sociopolitical, and cultural consequences of the AI's conceptual issues, technical imperfections, and biases; Żylińska's *AI Art* (2020) that discusses the AI's influences on the arts and culture; Crawford's *Atlas of AI* (2021) that maps the exploitative layers of AI research and business, hidden behind marketing, media hype, application interfaces, and cultural commodification; and Zeilinger's *Tactical Entanglements* (2021) whose multifaceted theoretical analysis of selected AI artworks focuses on their critical values, issues, and potentials.

Based on the earlier exploration of AI art (Grba 2020, 2021, 2022), this paper aims to diversify the existing discourse by introducing new critical perspectives on the poetic, expressive, and ethical features of tactical AI art. It describes diverse artistic approaches and their effectiveness in critiquing the phenomenological, epistemic, and political aspects of AI science and technology. The focus is on contextually relevant works that exemplify poetic complexity and manifest the epistemic or political ambiguities indicative of a broader milieu of contemporary art, culture, economy, and society.[1] This allows us to identify the conceptual, discursive, and ethical issues that affect the poetic outcomes and sociopolitical impact of tactical AI artworks, and to outline some of the prospects for the advancement of the field.

1. Footnotes list additional exemplars for further contextualization and comparison.

## 2. Subjects

Tactical AI art traces and challenges the constitution of social reality through the technical logic of AI that permeates the globalized infrastructures of industry, commerce, communication, entertainment, and surveillance. Artists uncover the problematic aspects and undesirable consequences of corporate AI and denounce biases, prejudices, economic inequalities, and political agendas encoded in the mainstream ML architectures. In some works, they also engage in an exploratory critique of the nature of ML as an artistic medium. To incite critical scrutiny, artists sometimes combine humour and provocation by intentionally taking ambivalent positions toward the issues they address; they emulate the corporate AI's operative models but recontextualize them or repurpose their objectives for ironic revelatory effects. One of the common methodologies involves taking an existent ML pipeline, training it with a non-standard dataset, and employing it for novel tasks. Successful works usually refrain from dramatic interventions and overly didactic explanations in order to let the audience actively identify the interests, animosities, struggles, inequalities, and injustices of corporate AI. A detailed study of tactical AI art would exceed the available volume of this paper, so its central discussion pertains to the exemplars in the three most representative subject areas: sociocultural, existential, and political.

### 2.1 Sociocultural

Many cultural manifestations of applied AI are linguistic, so artists often work with natural language processing (NLP) systems to critique their political undertones. For example, Jonas Eltes' installation *Lost in Computation* (2017) features a continuous real-time conversation between a Swedish-speaking and an Italian-speaking chatbot connected through Google Translate service. It simultaneously highlights the ambiguities of machine cognition and showcases the increasing accuracy and flexibility of language modelling algorithms (Eltes 2017). *Lost in Computation* references earlier Ken Feingold's installations such as *If, Then, What If,* or *Sinking Feeling* (all 2001) (Feingold 2021) and Marc Böhlen's *Amy and Klara* (2005-2008) (Böhlen 2005-2008). In these works, NLP systems provide semantically plausible but ultimately senseless continuation of narrative episodes which allude to the flimsiness of the Turing test and serve as vocalized metaphors for our lives. They extend the uncanny experience into the absurdity of miscommunication and accentuate the overall superficiality of the systems tasked to emulate human exchange. Artists also indicate the dubious sociopolitical background of NLP technologies. For example, Matt Richardson's *Descriptive Camera* (2012) has a lens but no display; it sends the photographed image directly to an Amazon MTurker tasked to write down and upload its brief description, which the device prints out (Richardson 2012). It provides a revelatory counter-intuitive glimpse into the widespread exploitation of transnational echelons of underpaid workers for ML training dataset annotation, which corporate AI euphemistically calls "artificial Artificial Intelligence" (AAI) or "pseudo-AI".[2]

To underline the issues in the visual layers of the AI-influenced culture, artists make deepfakes by modifying generative adversarial network (GAN) architectures. For example, Libby Heaney's *Resurrection (TOTB)* (2019) thematizes both the star power in music and the memetic power of deepfakes (Heaney 2019a). Visitors of this installation are invited to perform karaoke in which the original musician of a chosen song is video-deepfaked to mimic the visitor's singing and gesturing/dancing. In between karaoke acts, the host Sammy James Britten engages the audience in the discussion of power, desire, and control—an extension that seems to be as imposing and redundant as the artist's explanatory section for this work (Grba 2021, 246-247). Heaney's *Euro(re)vision* (2019) addresses the transmission of power and politics through popular media more effectively. In this video, deepfaked Angela Merkel and Theresa May sing absurd songs in the style of Dadaist Cabaret Voltaire performances within a setting of the Eurovision song contest (Heaney 2019b). Their disfluent algorithmic poetry eerily resembles the nonsensicality of actual Brexit discourse and implies the broader semantic reality of political life. In a similar fashion, Bill Posters and Daniel Howe confused the Instagram surfers by posting two iterations of their work *Big Dada: Public Faces* (2019-2021)—a series of deepfaked video statements by Marcel Duchamp about the ashes of Dada, Marina Abramović about mimetic evolution, Mark Zuckerberg about the second Enlightenment, Kim Kardashian about psycho-politics, Morgan Freeman about smart power, and Donald Trump about truth (Posters and Howe 2021).

In several works, Jake Elwes critically engages the cultural implications of training dataset annotation and algorithm design in mainstream AI. In the multi-part *Zizi Project* (since 2019), he interfaces deepfake with the world of LGBTQ+. *Zizi - Queering the Dataset* (2019) is a video installation continuously morphing through glitchy gender-fluid portraits. Elwes used a StyleGAN trained on Nvidia's Flickr-Faces-HQ dataset and retrained it on a new dataset of around 1,000 portraits of drag performers, scraped from the Internet. Another part of the *Zizi Project* is the online work *Zizi Show* (2020) which critiques both anthropomorphism and the error-prone gender inclusiveness of AI. This virtual drag cabaret features deepfakes generated from the training datasets based on the original films of London drag artists' performances (Elwes 2020). The *Zizi Project* clearly indicates that the training model datasets and statistical nature of data processing in GANs inevitably impose formal constraints on the possible outputs (such as realistic human-like images) regardless of the common rhetoric about the "unpredictability" or "originality" of such systems; however, this is an already known and well-documented issue (Pasquinelli 2019, 9-10). Beyond that, the project fails to show how exactly the race, gender, and class inequalities and stereotypes transfer into ML to harm the underrepresented social, ethnic, or gender groups. Its playful, technically sophisticated remediation within AI-influenced cultural context may be beneficial for the celebration, affirmation, and inclusion of LGBTQ+, but its publicity narratives, high production values, and focus on glamour and spectacle in lieu of less picturesque but perhaps more important existential aspects of LGBTQ+ can be perceived as exploitative. Moreover, if taken seriously by corporate AI, this critique can contribute to the refined normalization, instead of correction, of sociopolitical biases toward the LGBTQ+ community because these biases have a broader, deeper, and darker historical background.

---

[2]. Further examples of critical NLP include Ross Goodwin's *Text Clock* (2014); Michel Erler's *Deep Learning Kubrick* (2016); Ross Goodwin and Oscar Sharp's *Sunspring* (2016); Jonas Lund's *Talk to Me* (2017-2019); Joel Swanson's *Codependent Algorithms* (2018) (Swanson 2018); Disnovation. org's *Predictive Art Bot* (since 2017); Sofian Audry and Monty Cantsin's *The Sense of Neoism!* (2018); Philipp Schmitt's *Computed Curation Generator* (2017); Alexander Reben's *AI Am I (The New Aesthetic)* (2020); Nirav Beni's *AI Spy* (2020); and others.

     In contrast, Derek Curry and Jennifer Gradecki's deepfakes *Infodemic* (2020) and *Going Viral* (2020-2021) exemplify a consistently effective critique, recontextualization, and transformation of ML as a sociotechnical realm (Gradecki and Curry 2017). Both works target celebrities, pundits, politicians, and tech moguls who have "contributed" to the CoVID-19 pandemic by spreading misinformation and conspiracy theories, which themselves went "viral", often spreading faster than real news (Curry and Gradecki 2020; Gradecki and Curry 2021). For example, *Infodemic* features a cGAN-deepfaked talking head video in which some of these influencers deliver public service announcements voiced by academics, medical experts, and journalists that correct false narratives about the pandemic. By playing with deepfakes within their native context of fake news, these projects also probe the broader phenomenology of mediated narratives. The effectiveness of Curry and Gradecki's tactics is based on thorough research and self-referential critical methodology with computational media affordances; its playful transgressive affects are also friendly implications of our complicity to the politically problematic aspects of the applied AI through conformity, lack of involvement, or non-action. Leonardo Selvaggio's web project *Apologize to America* (2021) relates to this approach by using augmented reality instead of deepfake. Powered by Selvaggio's custom Snapchat lens, it invites visitors to record an apology while "wearing" the 45[th] President of the United States' por-

trait mapped onto their face. Recorded apologies are published and archived on the website apologize2america.com, and visitors can share them on social media (Selvaggio 2021).

### 2.2 Existential

AI technologies affect society, culture, and politics through the material, physical, ecological, and existential changes. Artists sometimes metaphorize this influence by using geospatial contents (landscapes, terrains, maps) for training datasets and by positioning the machine-learned output in politically connoted contexts.

For example, in several formally economical interactive installations, Nao Tokui addresses the arbitrariness of ML-powered sound and image recognition and synthesis in entertainment, advertising, surveillance, law enforcement, and the military. In *Imaginary Landscape* (2018), the software continuously analyses Google StreetView photographs, selects three that look similar, and joins them together horizontally in a three-wall projection. Another module, trained on landscape videos, generates soundscapes that correspond with stitched triptych landscapes (Tokui 2018a). In *Imaginary Soundwalk* (2018) viewers freely navigate Google StreetView for which the ML system, using the cross-modal technique for image-to-audio information retrieval, generates the "appropriate" soundscape (Tokui 2018b).

Some works explore the physicality of AI through haptics or kinesthetics. For example, François Quévillon's *Algorithmic Drive* (2018-2019) uses kinesthetics to play out the tension between robotics and the unpredictable nature of the world. For this work, several months-worth of data collected by a car's onboard computer, such as geolocation, orientation, speed, engine RPM, stability, and temperatures at various sensors, is synchronized with the video capture from the car's dash-cam. The captured videos and data feed a sampling system that sorts the content statistically and assembles a video that alternates between calm and agitated states by modifying the parameters of sound, image, car's activity, and environment. An interactive controller displays data for each scene and allows visitor intervention (Quévillon 2019).

Continuing the line of statistically driven eco-conscious works, such as Chris Jordan's *Running the Numbers* (since 2006) (Jordan 2021), artists use ML to generate visuals, objects, and narratives that address the environmental changes imposed by the large-scale computation-intense technologies of AI research and business. For example, Ben Snell's *Inheritance* (2020) elegantly compresses the material and ecological aspects of AI. It is a series of AI-generated sculptures cast in the composite medium which was produced by pulverizing the computers used to generate the sculptures' 3D models (Snell 2020). Although it is debatable how successfully this work deals with non-human agency and creative expression (Zeilinger 2021, 19-20),

64

it provocatively references radical auto-recursive art experiments such as Jean Tinguely's self-destructive machines. On the other hand, Maja Petrić's *Lost Skies* (2017) illustrates how easy it is for the projects in this range to slip into aestheticizing the ecological issues instead of articulating the data into meaningful or actionable narratives (Petrić 2021). Regardless of their poetic values, it is not easy to calculate, but probably not difficult to guess, the degree to which the systemic technological entanglements of eco-critical AI artworks (and AI art in general) themselves participate in the overall environmental damage.[3]

> [3]. Other examples include Tivon Rice's *Models for Environmental Literacy* (2020); Tega Brain, Julian Oliver, Bengt Sjölén's *Asunder* (2021); Kai-Luen Liang's *Blue Marbles* (2021); and others.

Max Hawkins' R*andomized Living* (2015-2017) features a more responsible integration and interrogation of a spectrum of the applied AI's material consequences. In this two-year experiment, Hawkins organized his life according to the dictate of recommendation algorithms. He designed a series of apps that used online data to suggest a city where he would live for about a month and, once there, the places to go, people to meet, and things to do (Hawkins 2021). *Randomized Living* is a bold exemplar of cybernetic existentialism—the art of conceiving a responsive and evolving cybernetic system in order to express deep existential concerns. Its implications involve the humans' general susceptibility to modifying behaviour and cognition in order to fit various machinic protocols, for example labour regimes in industrial capitalism or perceptual and interaction conditioning in early VR development (Lanier 2017). This susceptibility now manifests in a tendency among the users of AI-powered devices and the operators of AI systems to constrain their vocabulary and pronunciation so the software can interpret them (Pasquinelli 2019, 17). This reductivism is related to the shifts in social relations driven by the mutually reinforcing opportunism and network effects for the users of social media. It reflects the underlying pathological business logic of dominant information services, which dehumanize users and turn them into slavish data-generating commodities by addicting them to negatively biased, politically derisive, and socially toxic "free" services. It is worth noting, however, that such deviations are usually compensated by quick cultural maturation, as exemplified by the disproportionally high fidelity attributed by the audience to early photography, cinema, or sound recording whereas they later become aware of the artificiality and imperfections of these media. Nevertheless, while the specific AI issues can be viewed as transient side-effects of the continuing coevolution between culture and technology, it is important to remain cognizant and vigilant about them.

### 2.3 Political

In order to reverse-engineer the uneasy positioning of the individual toward or within computational systems of control, artists such as Josh On (On 2001-2004), Joana Moll (Moll 2020), Adam Harvey (Harvey and LaPlace 2021), and Vladan

4. Pioneered by Mark Lombardi in the 1990s and Bureau d'Etudes since the early 2000s, tactical cartography involves constructing diagrams and maps of financial and political power networks, which are simultaneously aesthetic, investigative, and activist (Hobbs and Richards 2003; Bureau d'Etudes 2015).

Joler have been using analytical tools and tactical cartography.[4] For example, with SHARE Lab and Kate Crawford, Vladan Joler released *Exploitation Forensics* (2017) whose series of intricate diagrams snapshots the functional logic of Internet infrastructure: from network topologies and the architecture of social media (Facebook) to the production, consumption, and revenue generation complex on Amazon.com (Anonymous 2017). Similarly, Joler and Crawford's collaborative project, *Anatomy of an AI System* (2018) deconstructs the Amazon Echo device's black box by mapping its components onto the frameworks of global economy and ecology (Crawford and Joler 2018). With Matteo Pasquinelli, Joler issued *The Nooscope Manifested* (2020), a visual essay about the structural and functional logic of subsymbolic ML, its epistemological and political implications (Joler and Pasquinelli 2020). It leverages the notions of gaze and vision-enhancing instruments as metaphorical and comparative devices, although their conceptual suitability within the context of ML is debatable.

Since the introduction of the OpenCV library in 2000, artists have been using computer vision (CV) for various purposes in a diverse corpus of exploratory works.[5] With advances in ML, this exploration has intensified and increasingly involved the critique of the (ab)use of CV for taxonomic imaging, object detection, face recognition, and emotion classification. For example, Jake Elwes' video *Machine Learning Porn* (2016) indicates human (perceptive) prejudices that influence the design of ML filters for "inappropriate" content. Elwes took the open_nsfw CNN that was originally trained with Yahoo's model for detecting "sexually explicit" or "offensive" visuals and repurposed its recognition classifiers as parameters for generating new images. This inversion outputs visually abstract videos with a "porny" allusiveness (Elwes 2016). However, the cogency of this project depends on leaving out that *all* visual forms are abstract by default and that the pathways of complex scene recognition and related decision-making in humans are not precisely known (Wang and Cottrell 2017; Wischnewski and Peelen 2021) so the ground for critiquing biases in these pathways is uncertain.

The sensitive issues of ML-powered biometry are particularly pertinent in facial recognition and classification due to the convergence of evolutionarily important visual features within the face and the psycho-social role of a face as the main representation of the self and identity. Various deficiencies frame the CV training and recognition processes in which the classification models ultimately always make implicit (but unobjective) claims to represent their subjects. The deficiencies in machinic face detection/identification, some of which have persisted from the earlier technologies such as VR (Lanier 2019), have been continuously identified by both scientists (Orcutt 2016; Zhao et al. 2017) and artists. For example, Joy Buolamwini and Timnit Gebru's scientific research, which started as Buolamwini's MS thesis in 2017, turned into a project with artistic overtones, titled *Gender Shades* (2018). It assesses the accuracy of several corporate facial classifiers (Adience, IBM, Microsoft, and Face++) with respect to gender, skin type, and skin type/gender intersection. Using a custom benchmark dataset with diverse skin types based on 1,270 images of parliamentarians from three African and three European countries, Buolamwini and Gebru showed that the error rate of the tested corporate classifiers was significantly higher for women with darker skin colour (Buolamwini and Gebru 2018). Encouraged by the IBM, they published their benchmark dataset so it can be applied in practice for accuracy calibration. Their findings affected the public, the corporate AI sector, and the US policymakers (Gershgorn 2020).

Kate Crawford and Trevor Paglen's exhibition *Training Humans* (2019-2020), and the accompanying essay *Excavating AI: The Politics of Images in Machine Learning Training Sets*, had a similar agenda (Crawford and Paglen 2019; Fondazione Prada 2020). Their critique of corporate practices for training CV systems includes racial bias and the use of facial images and videos without consent to build training datasets. Yet Michael Lyons, a co-author on one of the datasets featured in the project (JAFFE), showed that Paglen and Crawford themselves reproduced and exhibited these same images without consent (Lyons 2020, 2021; Leibowicz et al. 2021, 7). Compared with the methodologically flawed and ethically compromised strategy of *Training Humans*, Buolamwini and Gebru's *Gender Shades* similarly draws public attention to an established repertoire of race-related issues in CV, but it also productively intervenes in the tech and policy-making sectors, where such correctives (should) matter most. On the other hand, both of these projects open questions beyond the obvious technopolitical layers. Should an activist intervention end up (proactively or indirectly) improving the AI's profitable and further manipulable codes, and be used by the corporate sector to remedy its public image but without necessarily improving its technical and ethical standards? Or should it disrupt the code-crystalized corporate AI practices on a higher politically consequential level? And, within that context, how effectively the government policy changes can affect the private businesses with global influence?

Various modes of CV-driven interactivity (human-machine, machine-machine, and human-machine-human) allow artists to stir up a space for the audience's contemplation and critical interpretation. For example, Ross Goodwin's *word.camera* (2015) reiterates Matt Richardson's lexicographic approach in *Descriptive Camera* but uses "non-artificial" AI for image-to-text conversion.[6] The first version of this work prints out the passages from novels relevant to the uploaded images captured through a hand-held camera interface. The second version is a surveillance camera that autonomously searches for faces and describes them in "spoken" words (Goodwyn 2015). Jake Elwes' video installation *Closed Loop* (2017) establishes a mutually generative relational loop between a text-to-image and image-to-text model, whose inaccuracies and biases imply ethical issues in an unpredictable and witty continuum (Elwes 2017).

6. RyBN and Marie Lechner's media archeology project *Human Computers* (2016-2019) also uncovers the essential but largely "transparent" human echelons behind corporate AI.

Shinseungback Kimyonghun's installation *Mind* (2019) uses emotion analysis of the last 100 visitors' facial expressions to drive the ocean drums and generate a powerful minimalist sound ambient, with an overhead camera as a single indicator of the machinic gaze (Shinseungback Kimyonghun 2019). Martin Disley's open-source project *How They Met Themselves* (2021) exploits the recognition borders of face generation/recognition GANs. In a series of steps, it allows users to create photorealistic avatars for live webcam deepfaking. Based on the user's uploaded portrait, the avatar is created by a generation/discrimination process that yields two visually indistinguishable (virtually identical) images: one is positively identified as a person in the uploaded photo, and the other one is identified negatively (not a person in the photo). The user can then upload the generated ambivalent image to train the free online app Avatarify for real-time animated avatar superimposition in online interactions (Disley 2021).

Ironically, unlike the biases in ML, the individual "creative biases" and idiosyncrasies in AI art are desirable but relatively rare. Sebastian Schmieg tackles this deficiency with conceptual relevance, expressive economy, and formal clarity in projects such as *Decision Space* (2016); *This is the Problem, the Solution, the Past and the Future* (2017); *Decisive Camera* (2017-2018); and *Decisive Mirror* (2019) (Schmieg 2016, 2017, 2018, 2019). In different ways, these works inject unconventional, seemingly absurd, or counter-intuitive taxonomies into image classification setups. For example, the visitors of the *Decisive Camera* project website can upload an image which will then be classified within a taxonomic space of four categories: Problem, Solution, Past, and Future, and assigned with a probability percentage for each category. The classification dataset was created in the project's initial phase which invited visitors to select images from the Photographers Gallery's image archive and to assign each image to one of these four categories. This playful subversion places the technical, methodological, and broader sociopolitical problems of ML design conventions firmly within the human context. It also provides the reflections of human nature in the arbitrary authoritarianism of corporate ML classification systems based on exploiting human labour for annotating the training datasets.

Artists also critique the human appetite for exploiting the speculative investment strategies wetted by corporate AI and related crypto technologies. For example, Anna Ridler's *Mosaic Virus: Bitcoin Per Hour* (2018) questions the concepts of ownership, obsessions with wealth, and financial speculation by referring to the historical "tulip mania" phenomenon. Trained on Ridler's custom dataset of hand-labelled photographs of tulips, a GAN generates images of tulips inflected by the current Bitcoin values. It links the instability of values projected onto commodified artefacts with the opacity of computational technologies used in creating the work (Ridler 2019). Ben Bogard's *Zombie Formalist* (2021) arranges a witty marriage of ML and Komar and Melamid's *People's Choice* (1994-1997)[7] aiming to direct a critical focus onto the hyperproduction of bland formalism (Robinson 2014) as a signifier of digital art's commodification boosted by the crypto art market. In this installation, two AI-powered lightboxes randomly generate abstract images calibrated by measuring the viewers' engagement in two modes: the attention span via face detection, and the number of Twitter likes and retweets of the uploaded images (Bogart 2021a). Both *Mosaic Virus* and *Zombie Formalist* make clear cases, but mainly for the audience that is already critically aware of Bitcoin politics or digital art's commodification. For the average audience—which may be unfamiliar with their specific issues—the strong aesthetic fronts of these projects can be decisive or counter-effective. As is often the case with tactical art, the combination of lofty motivation and somewhat ambiguous presentation may diminish the projects' effectiveness or even expedite its recuperation. Since the sociotechnical unpredictability is closely related to financial instability, it is worth remembering that AI research, which has been going through successive "springs" and "winters" (Mitchell 2019, 31-32), may end up in Disnovation.org's project *The Museum of Failures* (since 2015). In a museological setting, it features a collection of aborted tech projects, flops, errors, malfunctions, business failures, ethical rejections, or disasters presented in various formats from historical, symbolic, poetic, and cultural points of view (Disnovation.org 2021).

---

[7]. Although Bogart does not acknowledge this referential work (Bogart 2021a; 2021b).

## 3. Challenges

These examples show that, through success or failure, tactical AI art reveals human fallacies, conceptual constraints, and sociopolitical ambiguities in both the AI-influenced society and in AI art itself. By identifying, acknowledging, and understanding these issues, artists can find new ways to intervene critically and productively in current sociopolitical reality. Similar to AI research's struggles with encoding crucial aspects of human cognition such as intuition, abstraction, analogy-making, common sense, and inventiveness into machine intelligence (Mitchell 2019, 200-214; Marcus and Davis 2019, 160-191), the poetic realm of AI art is deficient in interesting intuitions, meaningful abstractions, strong concepts, and imaginative analogies that effectively address the wider perspectives or deeper issues of human existence. The uneven intellectual breadth and depth, biased or constrained contextual awareness, and sketchy art-historical knowledge affect many artists' conceptual thinking, methodologies, and the cogency of their outcomes.

Technocratic or techno-fetishist mentalities have been haunting computational arts since their outset, and continue to affect AI art (Taylor 2014; Żylińska 2020, 75-85). They often reinforce a naïve lack of understanding that production techniques in the arts fundamentally unfold and get emancipated by coupling with conceptual thinking and contextual awareness. Conversely, artists who exaggerate or fake technical competencies are equally problematic because their works usually miss interesting technological aspects.

Various apparent, but often undisclosed conceptual, methodological, thematic, aesthetic, and presentational similarities between different works indicate the issues of the artists' creative literacy and contextual appreciation. Sincerely-motivated and well-conceived concepts are sometimes rendered as dry, unengaging, ineffective, or counter-effective works (Grba 2022, 20-21). Critical cogency, viability, and impact are affected by the pretentious or didactic representational strategies, exceeding topicality, and inflated theoretical rhetoric (Quaranta 2020; Grba 2021, 246-247). Furthermore, artistic and academic communities tend to develop echo chambers in which their work gains significance while its real-world impact requires more stringent assessment and correctives. The virtualization of critical focus (or purpose) may be fruitful within the academic milieu, but the general audience, which is central to tactical art, can easily recognize it as aloofness or cynicism which leads to indifference, distrust, or resentment.

Broader issues that affect tactical AI art include the uninformed media coverage, the questionable norms of the art community, the depleting autonomy of academic institutions, and the problematic legal norms for centralized, profit-motivated control of intellectual property and creative labour. The responsibility for tackling these issues lies not only with the artists, but also with scientists, entrepreneurs, cultural agents, and the public. The exploitative strategies of MCA entice AI artists to compromise their tactical goals in order to accommodate the conservative requirements for scarcity, commercial viability, and ownership (Grba 2020, 252-254). As they unfold within the bubblingly scammy NFT ecosystem (Quaranta 2021), the artists' proverbial inclinations toward myopic opportunism call for sophisticated tactical interventions that would disrupt the lures of commodification, complacency, and recuperation. In general, it is important to acknowledge that both art and technology are human dispositives within anthropological and sociocultural perspectives so that the poetic qualities of our artefacts are inherently instrumentalizable as virtue signalling means driven by competitive ambitions.

## 4. Perspectives

Contemporary AI provides an excellent milieu for the artists to demystify ML systems as sociopolitical apparatuses and to reiterate that science, technology, and businesses need thorough improvement of epistemological and ethical standards facing the increasing complexity of human existence. Therein lays the potential for tactical AI art to direct computational arts toward a socially responsible and epistemologically relevant expressive stratum.

In order to engage the audience with a lasting impact, artists need to match their procedural skills with motivational sincerity and ideational cogency, and maintain a critical outlook on their poetic devices. The ethos of maturely balanced competencies deserves cultivation through expressive diversification, experimental freedom, playfulness, bricolage, conceptually strong hacking, and imaginative discovery. In principle, artists can benefit from epistemic humility to develop more rigorous criteria for creative thinking and better multidisciplinary knowledge of historical, theoretical, cultural, and political contexts in which they produce and present their works (Böhlen 2020; Grba 2022, 24). This will help them address the potentially adverse scenarios and clear the way for meaningful creative directives. The inherently political nature of AI (Pasquinelli 2019) obliges artists not only to exploit but to deconstruct and explore their expressive means by recognizing the injustices in the notional, relational, technical, social, and other layers of the conditions in which they live and create.

However, all actors in AI art should strive for integrity by recognizing, objectively assessing, and correcting the systemically biased and noisy professional value systems of the MCA and academia. The entanglements with corporate AI, MCA, and academia support the forthcoming AI art projects, but may also attenuate their criticality and expedite recuperation. To tackle this, artists should strive to resist prioritizing their careers over their art, be open to taking genuine risks by evolving potentially hazardous ideas, and pursue systematic support with scepticism toward institutional rationales for art sponsorship. Successful tactical projects utilize their entanglements self-consciously, as the conceptual and existentially inherent features of digital culture. Their impact can be improved by bolder and more nuanced examination of the cultural and sociopolitical contexts of AI technology and business, and by deeper probing and problematizing the underlining concepts such as intelligence, creativity, expressive agency, authorship, intellectual labour, ownership, authenticity, accuracy, and bias.

For countering the recuperative sophistication of info-capitalism (which artists tend to underestimate, overlook, or ignore), stealthy subversiveness and subterfuge seem to be more prudent than didactic overexplanation or overbearing spectacularism. Artists should articulate and respect their methodologies as heterogeneous productive frameworks whose experiential processes and outcomes inform the audience by stirring inquisitiveness and critical thinking, stimulating imagination, and encouraging progressive action. Such tactical frameworks are more impactful than surface-based, aestheticized, descriptive, or rhetorical ones. By demystifying the seemingly radical capabilities of their tools, artists can leverage the issues of modern AI as critical assets with wide political significance. Empowered by the destabilizing value of humour, responsible treatment of these assets can build new insights about human nature and provide meaningful posthumanist perspectives (McQuillan 2018).

# References

**Anonymous.**
2017. "SHARE Lab: Exploitation Forensics Press Release." Aksioma – Institute for Contemporary Art website. http://www.aksioma.org/press/exploitation.forensics.zip

**Barbrook, Richard, and Andy Cameron.**
1995/2008. "The Californian Ideology." *Mute*, (1), 3, Code. http://www.metamute.org/editorial/articles/californian-ideology

**Bishop, Claire.**
2012. "Digital Divide: Contemporary Art and New Media." *Artforum*, 51, 1: 434–441.

**Bogart, Ben.**
2021a. "Zombie Formalist 2021–." Ben Bogart's website. https://www.ekran.org/ben/portfolio/2021/07/the-zombie-formalist-2021

**Bogart, Ben.**
2021b. "The Zombie Formalist: An Art Generator that Learns." *Art Machines 2: International Symposium on Machine Learning and Art Proceedings.* School of Creative Media, City University of Hong Kong: 165-166.

**Böhlen, Marc.**
2005-2008. "Amy and Klara." Marc Böhlen's website. https://realtechsupport.org/projects/maledicta.html

**Böhlen, Marc.**
2020. "AI Has a Rocket Problem." Medium, 3 August. https://medium.com/swlh/ai-has-a-rocket-problem-6949c6ed51e8

**Bourriaud, Nicolas.**
2002. *Relational Aesthetics* (translated by Simon Pleasance and Fronza Woods), 14. Paris: Les Presses du Réel.

**Buolamwini, Joy, and Timnit Gebru.**
2018. "Gender Shades: Intersectional Accuracy Disparities in Commercial Gender Classification." *Proceedings of the 1st Conference on Fairness, Accountability and Transparency 81:* 77-91. https://proceedings.mlr.press/v81/buolamwini18a.html

**Burbano, Andrés, and Ruth West, eds.**
2020. "AI, Arts & Design: Questioning Learning Machines." *Artnodes*, (26): 1-9 (2-3). http://doi.org/10.7238/a.v0i26.3390

**Bureau d'Etudes (Léonore Bonaccini and Xavier Fourt).**
2015. "Bureau d'Etudes." Bureau d'Etudes' blog. https://bureaudetudes.org

**Crawford, Kate, and Trevor Paglen.**
2019. "Excavating AI: The Politics of Training Sets for Machine Learning." Project website. https://excavating.ai

**Crawford, Kate, and Vladan Joler.**
2018. "Anatomy of an AI System." Project website. https://anatomyof.ai

**Crawford, Kate.**
2021. *Atlas of AI: Power, Politics, and the Planetary Costs of Artificial Intelligence.* New Haven and London: Yale University Press.

**Curry, Derek, and Jennifer Gradecki.**
2020. "Infodemic." Derek Curry's website. https://derekcurry.com/projects/infodemic.html

**Disley, Martin.**
2021. "How They Met Themselves." Martin Disley's website. http://www.martindisley.co.uk/how-they-met-themselves

**Disnovation.org (Nicolas Maigret and Maria Roszkowska).**
2021. "The Museum of Failures." Project website. https://disnovation.org/mof.php

**Eltes, Jonas.**
2017. "Lost in Computation." Jonas Eltes' website. https://jonaselt.es/projects/lost-in-computation

**Elwes, Jake.**
2016. "Machine Learning Porn." Jake Elwes' website. https://www.jakeelwes.com/project-mlporn.html

**Elwes, Jake.**
2017. "Closed Loop." Jake Elwes' website. https://www.jakeelwes.com/project-closedLoop.html

**Elwes, Jake.**
2020. "The Zizi Show." Jake Elwes' website. https://www.jakeelwes.com/project-zizi-show.html

**Feingold, Ken.**
2021. "Works." Ken Feingold's website. https://www.kenfeingold.com

**Fondazione Prada.**
2020. "Kate Crawford | Trevor Paglen: Training Humans." Fondazione Prada website. https://www.fondazioneprada.org/project/training-humans

**Gershgorn, Dave.**
2020. "How a 2018 Research Paper Led Amazon, Microsoft, and IBM to Curb Their Facial Recognition Programs." OneZero Medium website. https://onezero.medium.com/how-a-2018-research-paper-led-to-amazon-and-ibm-curbing-their-facial-recognition-programs-db9d6cb8a420

**Goodwin, Ross.**
2015. "Narrated Reality." Ross Goodwin's website. https://rossgoodwin.com/narratedreality

**Gradecki, Jennifer, and Derek Curry.**
2017. "Crowd-Sourced Intelligence Agency: Prototyping Counterveillance." *Big Data & Society*, January–June: 1–7. https://journals.sagepub.com/doi/

**Gradecki, Jennifer, and Derek Curry.**
2021. "Going Viral." Project website. https://goingviral.art

**Grba, Dejan.**
2020. "Alpha Version, Delta Signature: Cognitive Aspects of Artefactual Creativity." *Journal of Science and Technology of the Arts*, 12 (3): 63-83. https://doi.org/10.34632/jsta.2020.9491

**Grba, Dejan.**
2021. "Brittle Opacity: Ambiguities of the Creative AI." In M. Verdicchio, M. Carvalhais, L. RIbas, A. Rangel (editors), Proceedings of xCoAx 2021, 9th Conference on Computation, Communication, Aesthetics & X. Porto: 235-260. ISBN: 978-989-9049-06-2. https://doi.org/10.5281/zenodo.5831884

**Grba, Dejan.**
2022. "Deep Else: A Critical Framework for AI Art." *MDPI Digital*, (2), 1: 1-32. ISSN 2673-6470. https://doi.org/10.3390/digital2010001

**Harvey, Adam, and Jules LaPlace.**
2021. "Exposing.ai." Project website. https://exposing.ai

**Hawkins, Max.**
2021. "Randomized Living." Max Hawkins' website. https://maxhawkins.me/work/randomized_living

**Heaney, Libby.**
2019a. "Resurrection (TOTB)." Libby Heaney's website. http://libbyheaney.co.uk/resurrection-totb

**Heaney, Libby.**
2019b. "Euro(re)vision." AI Art Gallery website. http://www.aiartonline.com/highlights/libby-heaney-2

**Hobbs, Robert Carleton, and Judith Richards**.
2003. *Mark Lombardi: Global Networks.* Independent Curators International. ISBN 0-916365-67-0.

**Joler, Vladan, and Matteo Pasquinelli.**
2020. "The Nooscope Manifested." Project website. https://nooscope.ai

**Jordan, Chris.**
2021. "Running the Numbers." Chris Jordan's website. http://chrisjordan.com/gallery/rtn

**Kearns, Michael, and Aaron Roth.**
2019. *The Ethical Algorithm: The Science of Socially Aware Algorithm Design.* Oxford: Oxford University Press.

**Lanier, Jaron.**
2017. *Dawn of the New Everything*, 143. New York: Henry Holt and Com.

**Lanier, Jaron.**
2019. "Gray Area Festival Keynote." (02:35 - 03:00). Gray Area YouTube channel. https://youtu.be/lsNF4KfmwkY

**Leibowicz, Claire, Emily Saltz, and Lia Coleman.**
2021. "Creating AI Art Responsibly: A Field Guide for Artists." *Diseña*, 19 (September), Article 5. https://doi.org/10.7764/disena.19.Article.5. https://www.partnershiponai.org/wp-content/uploads/2020/09/Partnership-on-AI-AI-Art-Field-Guide.pdf

**Lovink, Geert, and Ned Rossiter.**
2005. "Dawn of the Organised Networks." Nettime, (October 17).

**Lyons, Michael J.**
2020. "Excavating 'Excavating AI': The Elephant in the Gallery." ACM ArXiv Preprint, September, (arXiv:2009.01215). https://doi.org/10.5281/zenodo.4391458

**Lyons, Michael J.**
2021. "'Excavating AI' Re-excavated: Debunking a Fallacious Account of the JAFFE Dataset." Preprint, 28 July. https://doi.org/10.5281/zenodo.5140556 / https://arxiv.org/abs/2107.13998

**Marcus, Gary F., and Ernest Davis.**
2019. *Rebooting AI: Building Artificial Intelligence We Can Trust.* New York: Pantheon Books.

**McQuillan, Dan.**
2018. "Manifesto on Algorithmic Humanitarianism." *Reimagining Digital Humanitarianism.* London: Goldsmiths, University of London. https://www.researchgate.net/publication/323689097_The_Manifesto_on_Algorithmic_Humanitarianism

**Mitchell, Melanie.**
2019. *Artificial Intelligence: A Guide for Thinking Humans.* Kindle edition. New York: Farrar, Straus and Giroux.

**Moll, Joana.**
2020. "Joana Moll." Joana Moll's website. http://www.janavirgin.com/index.html

**On, Josh.**
2001-2004. "They Rule." Josh On's YouTube channel. https://youtu.be/DPLD8uHxdtg

**Orcutt, Mike.**
2016. "Are Face Recognition Systems Accurate? Depends on Your Race." MIT Technology Review website, 6 July. https://www.technologyreview.com/2016/07/06/158971/are-face-recognition-systems-accurate-depends-on-your-race

**Pasquinelli, Matteo.**
2019. "How a Machine Learns and Fails—A Grammar of Error for Artificial Intelligence." *Spheres—Journal for Digital Cultures*, 5: 1-17. https://spheres-journal.org/wp-content/uploads/spheres-5_Pasquinelli.pdf

**Petrić, Maja.**
2017. "Lost Skies." Maja Petrić's website. https://www.majapetric.com/lost-skies

**Posters, Bill, and Daniel Howe.**
2021. "Big Dada: Public Faces." Bill Posters' website. http://billposters.ch/projects/big-dada

**Quaranta, Domenico.**
2020. "Between Hype Cycles and the Present Shock: Art at the End of the Future." *DOC#6*. Nero Publications. https://www.neroeditions.com/docs/between-hype-cycles-and-the-present-shock

**Quaranta, Domenico.**
2021. *Surfing con Satoshi: Arte, blockchain e NFT.* Milano: Postmedia Books.

**Quévillon, François.**
2019. "Algorithmic Drive." François Quévillon's website. http://francois-quevillon.com/w/?p=1466

**Raley, Rita.**
2009. *Tactical Media*, 1-30. Minneapolis: University of Minnesota Press.

**Richardson, Matt.**
2012. "Descriptive Camera™." Matt Richardson's website. http://mattrichardson.com/Descriptive-Camera/index.html

**Ridler, Anna.**
2019. "Mosaic Virus." Anna Ridler's website. http://annaridler.com/mosaic-virus

**Robinson, Walter.**
2014. "Flipping and the Rise of Zombie Formalism." Artspace. https://www.artspace.com/magazine/contributors/see_here/the_rise_of_zombie_formalism-52184

**Schmieg, Sebastian.**
2016. "Decision Space." Sebastian Schmieg's website. https://sebastianschmieg.com/decision-space

**Schmieg, Sebastian.**
2017. "This is the Problem, the Solution, the Past and the Future." Sebastian Schmieg's website. https://sebastianschmieg.com/this-is-the-problem-the-solution-the-past-and-the-future

**Schmieg, Sebastian.**
2018. "Decisive Camera." Sebastian Schmieg's website. https://sebastianschmieg.com/decisive-camera

**Schmieg, Sebastian.**
2019. "Decisive Mirror." Sebastian Schmieg's website. https://sebastianschmieg.com/decisive-mirror

**Selvaggio, Leonardo.**
2021. "Apologize to America." Leonardo Selvaggio's website. http://leoselvaggio.com/new-page

**Shinseungback Kimyonghun.**
2019. "Mind." Shinseungback Kimyonghun's website. http://ssbkyh.com/works/mind

**Snell, Ben.**
2020. "Inheritance." Ben Snell's website. http://bensnell.io/inheritance-ii

**Taylor, Grant D.**
2014. *When the Machine Made Art: The Troubled History of Computer Art.* New York and London: Bloomsbury Press.

**Tokui, Nao.**
2018a. "Imaginary Landscape." Nao Tokui's website. https://naotokui.net/projects/imaginary-landscape-2018

**Tokui, Nao.**
2018b. "Imaginary Soundwalk." Nao Tokui's website. https://naotokui.net/projects/imaginary-soundwalk-2018

**Wang, Panqu, and Garrison W. Cottrell.**
2017. "Central and Peripheral Vision for Scene Recognition: A Neurocomputational Modeling Exploration." *Journal of Vision*, 17, (4):9: 1-22. doi:10.1167/17.4.9.

**Wilson, Stephen.**
2002. *Information Arts: Intersections of Art, Science and Technology*, 786-811. Cambridge: MIT Press.

**Wischnewski, Miles, and Marius V. Peelen.**
2021. "Causal Neural Mechanisms of Context-Based Object Recognition." *eLife*, 10, e69736.

**Zeilinger, Martin.**
2021. *Tactical Entanglements: AI Art, Creative Agency, and the Limits of Intellectual Property.* Lüneburg: meson press.

**Zhao, Jieyu, Tianlu Wang, Mark Yatskar, Vicente Ordonez, and Kai-Wei Chang.**
2017. "Men Also Like Shopping: Reducing Gender Bias Amplification Using Corpus-Level Constraints." *Proceedings of the 2017 Conference on Empirical Methods in Natural Language Processing:* 2979-2989. https://doi.org/10.18653/v1/D17-1323

**Żylińska, Joanna.**
2020. *AI Art: Machine Visions and Warped Dreams.* London: Open Humanities Press.